\documentclass{mn2e}
\usepackage{epsfig}

\newcommand{\mnref}[1]{\hangindent=0.5in \hangafter=1 #1 \par}
\newenvironment{refs}{\parindent=0pt}{\parindent=1.5em}
\newcommand{\mn}{MNRAS}
\newcommand{\mnras}{MNRAS}

\newcommand{\aj}{AJ}
\newcommand{\apj}{ApJ}
\newcommand{\apjs}{ApJS}
\newcommand{\aaa}{A\&A}
\newcommand{\aap}{A\&A}
\newcommand{\aas}{A\&AS}

\newcommand{\Msolar}{\mbox{\,$\rm M_{\odot}$}}
\newcommand{\Lsolar}{\mbox{\,$\rm L_{\odot}$}}
\def\gs{\mathrel{\raise1.16pt\hbox{$>$}\kern-7.0pt
\lower3.06pt\hbox{{$\scriptstyle \sim$}}}}
\def\ls{\mathrel{\raise1.16pt\hbox{$<$}\kern-7.0pt
\lower3.06pt\hbox{{$\scriptstyle \sim$}}}}

\title{Helium and Hydrogen Line Ratios and 
The Stellar Content of Compact HII Regions} 
\author[S.L. Lumsden, P.J. Puxley, M.G. Hoare, T.J.T. Moore, N.A. Ridge]
{S.L. Lumsden$^{1}$, P.J. Puxley$^{2}$, M.G. Hoare$^{1}$, T.J.T. Moore$^{3}$
and N.A. Ridge$^{4}$\\
{}$^1$ {\em Department of Physics and Astronomy,
University of Leeds, Leeds LS2 9JT, UK}\\
{}$^2$ {\em Gemini Observatory, 670 N. A'ohuku Place, Hilo, Hawaii 96720, 
USA}\\
{}$^3$ {\em Astrophysics Research Institute, Liverpool John Moores University,
 Twelve Quays House, Egerton Wharf, Birkenhead CH41 1LD, UK}\\
{}$^4$ {\em Five College Radio Astronomy Observatory, 
619 Lederle Graduate Research Center, University of Massachusetts,
Amherst MA 01003}\\
{Email -- sll@ast.leeds.ac.uk, ppuxley@gemini.edu, mgh@ast.leeds.ac.uk,
tjtm@astro.livjm.ac.uk, naomi@fcrao1.astro.umass.edu}\\
}

\begin{document}

\label{firstpage}

\maketitle

\begin{abstract}
We present observations and models of the behaviour of the HI and HeI lines
between 1.6 and 2.2$\mu$m in a small sample of compact HII regions.  As in our
previous papers on planetary nebulae, we find that the `pure' 1.7007$\mu$m
4$^{3}$D--3$^{3}$P and 2.16475$\mu$m 7$^{3,1}$G--4$^{3,1}$F HeI recombination
lines behave approximately as expected as the effective temperature of the
central exciting star(s) increases.  However, the 2.058$\mu$m 2$^1$P--2$^1$S
HeI line does not behave as the model predicts, or as seen in planetary
nebulae.  Both models and planetary nebulae showed a decrease in the HeI
2$^1$P--2$^1$S/HI Br$\gamma$ ratio above an effective temperature of 40000K.
The compact HII regions do not show any such decrease.  The problem with this
line ratio is probably due to the fact that the photoionisation model does not
account correctly for the high densities seen in these HII regions, and that we
are therefore seeing more collisional excitation of the 2$^1$P level than the
model predicts.  It may also reflect some deeper problem in the assumed model
stellar atmospheres.  In any event, although the normal HeI recombination lines
can be used to place constraints on the temperature of the hottest star
present, the HeI 2$^1$P--2$^1$S/HI Br$\gamma$ ratio should not be used for this
purpose in either Galactic HII regions or in starburst galaxies, and
conclusions from previous work using this ratio should be regarded with extreme
caution.  We also show that the combination of the near infrared `pure'
recombination line ratios with mid-infrared forbidden line data provides a good
discriminant of the form of the far ultraviolet spectral energy distribution of
the exciting star(s).  From this we conclude that CoStar models are a poor
match to the available data for our sources, though the more recent WM-basic
models are a better fit.
\end{abstract}


\section{Introduction}

This is the last paper in a series discussing the near infrared hydrogen and
helium recombination lines in photoionised nebulae.  In the previous two papers
(Lumsden, Puxley \& Hoare 2001a,b; hereafter Paper 1 and 2) we discussed the
properties as measured in planetary nebulae.  We showed that the ratio of pure
hydrogen and helium recombination lines are a good measure of the hardness of
the exciting radiation field from the central star.  This is as expected from
simple photoionisation models, since the volume of the He$^+$ zone relative to
the volume of the H$^+$ zone is simply related to the hardness of the radiation
field in the 504--912\AA\ region of the spectrum of the exciting star.  In turn
this depends on the stellar effective temperature as well as more specific
details of the stellar spectrum as related to stellar evolution (ie stellar
winds, luminosity, surface gravity etc).  For young stars, this ratio is also
therefore clearly a constraint on the upper end of the stellar initial mass
function (IMF).

The use of near infrared hydrogen and helium lines as probes of the underlying
radiation field was suggested by Thompson \& Tokunaga (1980) in the study of
heavily embedded compact HII regions.  With early infrared instruments the near
infrared pure recombination HeI lines were too weak to be useful.  Considerable
attention was given instead to the HeI 2$^1$P--2$^1$S line at 2.058$\mu$m.
Unfortunately this line is largely pumped by the resonance in the HeI
2$^1$P--1$^1$S transition, and can be influenced by collisional excitation from
the metastable 2$^1$S level as well as from the triplet series.  At first it
was believed that the complexity of this transition could be modelled in a
simple fashion.  Doyon, Puxley \& Joseph (1992) suggested that the HeI
2$^1$P--2$^1$S to HI Br$\gamma$ ratio was ideal for studying the initial mass
function in starburst galaxies.  They tried to take account of the competing
factors that led to emission in the 2.058$\mu$m line using an analytical
approach.  However, Shields (1993) showed from a full photoionisation treatment
that this approach was a poor approximation to the actual predicted line
strength.  More recent updates to the atomic data used in the prediction
of the HeI 2$^1$P--2$^1$S line strength have also been presented by 
Ferland (1999).

In the previous two papers in this series we were able to show that the
observed near infrared HeI and HI line strengths did largely follow the
predictions of the photoionisation models for planetary nebulae.  This is
encouraging since the planetary nebulae are at least simple in the sense that
the objects we observed only had a single central exciting
star, and could be modelled reasonably as expanding spheres.  If the model
predictions for such simple systems deviated significantly from the
observations, then the method of using near infrared HeI and HI line strengths
as a constraint on the hardness of the stellar radiation field would have been
invalidated almost completely.  The aim of this paper is to study how well the
line ratios compare between model and observation for compact HII regions.

HII regions provide a more challenging test of these models for several
reasons.  They probably contain a central star cluster rather than a single
star, so the cumulative effect of the stellar output needs to be estimated.
This in turn requires knowledge of the IMF.  Of course, this is the parameter
that Doyon et al.\ (1992) sought to estimate from the HeI to HI line ratios.
However, in our case, where we are testing the models, we are forced to make
some simple assumptions regarding the form of the initial mass function.
Compact HII regions are also rather more diverse morphologically than the
simple planetary nebulae we were considering.  Finally, there is considerable
uncertainty about the form of the stellar radiation field in massive stars.
Models diverge markedly depending on whether line blanketing due to metals is
included or not, whether the effects of a stellar wind are included, and
whether a non-LTE treatment is applied.  Martins, Schaerer \& Hillier (2002)
give a useful summary of the properties of the currently available models
for dwarfs.  In
addition to this overall uncertainty, there is also some doubt as to whether
very young massive stars have the same effective radiation field as they would
if they were on the main sequence.  We know that massive young stellar objects
tend to have very dense, slow moving, stellar winds (eg.\ Bunn, Drew \& Hoare
1995).  These winds may reprocess much of the ionising radiation from the young
stars into non-ionising radiation.  The objects studied by Bunn et al.\ do not
have substantial HII regions despite having the bolometric luminosity
equivalent to OB stars.  The aim of this paper therefore is to determine how
well we can reproduce the HeI/HI line ratios used in Papers 1 and 2 using
realistic models of compact HII regions.

\section{The observational sample}
\subsection{Near infrared spectroscopy}
All of our target HII regions are drawn from the compilation of ultracompact
HII regions presented by Wood \& Churchwell (1989) and Kurtz, Churchwell \&
Wood (1994).  The objects were chosen to be sufficiently
bright in the near infrared to allow high signal-to-noise low resolution
spectroscopy to be obtained.
The sources
have a variety of morphologies, at least as seen in the radio continuum images.

The line ratios presented in this paper were taken from spectroscopic data
acquired for other purposes.  Most of the data were obtained using the facility
near infrared spectrograph CGS4 at the United Kingdom Infrared Telescope.  The
high spectral resolution data from which we draw the 2.16475$\mu$m HeI
7$^{3,1}$G--4$^{3,1}$F/HI Br$\gamma$ line ratio comes from a project to map the
kinematics of compact HII regions (eg.\ Lumsden \& Hoare 1996, 1999).  The
original data have a spectral resolution of $\sim20000$, sufficient to easily
distinguish the satellite HeI lines from Br$\gamma$ itself.  Full details of
the observations and data reduction can be found in the cited papers.  

The lower resolution data comes from a project to study the near infrared
extinction law (Ridge et al.\ in preparation).  Here observations of the
1.6--2.2$\mu$m region were acquired at low spectral resolution ($R\sim$400).
All of the data were acquired with CGS4 with the exception of that for
G45.45+0.06, where we have used data acquired with the near infrared
imager/spectrograph IRIS on the AAT.  The raw data were processed, corrected
for atmospheric absorption and flux calibrated in the usual fashion (see, eg,
Paper 2 which discusses data acquired on planetary nebulae using CGS4, or
Lumsden \& Puxley 1996 which deals with similar IRIS data).  We measured the
fluxes of the HI Br series lines, in order to determine the extinction, as well
as 2.058$\mu$m HeI 2$^1$P--2$^1$S and 1.7007$\mu$m HeI 4$^{3}$D--3$^{3}$P from
the low resolution data.  Unfortunately, the H band data for G45.07+0.13 are of
insufficient signal-to-noise to detect the HI Br12 or HeI 4$^{3}$D--3$^{3}$P
lines.  The limits that can be placed are consistent with other data
for this object, without actually adding any extra constraints.

For both high and low resolution data, we measured the line fluxes from the
spectrum summed over the whole object.  There was no evidence for a variation
in line ratio along the slit in any of our sources.  Since we are only
interested in line ratios with HI Br$\gamma$, we have combined the low and high
resolution data in our analysis.  The line ratios from the high resolution data
are not corrected for extinction. The small wavelength range involved implies a
correction of less than 1\% for all our sources.  The low resolution data were
corrected by comparing the HI lines of Br$\gamma$, Br10, Br11 and Br12 (the
latter 3 lines lie at 1.737$\mu$m, 1.681$\mu$m and 1.644$\mu$m).  These line
ratios vary by less than 5\% over the expected range of electron temperatures
and densities (eg Wood \& Churchwell 1989).  We assumed a power law dependence
for the extinction so that $\tau\propto\lambda^{-1.85}$ (Landini et al.\ 1984).
The error in the HeI 1.7007$\mu$m/HI Br$\gamma$ ratio is then largely set by
the derived error in this extinction correction.

The HeI 2$^1$P--2$^1$S line must also be corrected for a high frequency
component in the atmospheric absorption that is not removed by standard
techniques (see Paper 1 for full details).  The adopted correction values are
given to allow the raw line ratios to be estimated.  Table 1 presents these
values, all the observed line ratios after correction for extinction and the
adopted values of $\tau_{\rm{Br}\gamma}$.

Finally we also use data on the HeI 2$^1$P--2$^1$S/HI Br$\gamma$ ratio
previously published by Doherty et al.\ (1994) for G5.97--1.17 and
G43.89--0.78.  The extinction corrections applied there are derived from a
comparison of radio continuum and Br$\gamma$ flux density.  We give the values
as taken from the original source.

\subsection{Radio and Far Infrared Data}
We used the data given in Wood \& Churchwell (1989) and Kurtz et al.\ (1994)
in order to constrain the IMF for the exciting stars in these sources.  
Table 2 gives the basic observational properties of our sample.  
The
radio data gives a measure of the flux of hydrogen ionising photons, whereas
IRAS data gives a measure of the total luminosity.  The comparison of radio and
far infrared data therefore gives us information as to whether the exciting
source in an HII region is a single star or a cluster of stars since the two
observable properties depend on stellar mass (and the IMF) in a slightly
different way (see Section 3.2).

The radio data actually provides a lower limit to the flux of hydrogen ionising
photons.  The interferometric data presented in Wood \& Churchwell and Kurtz et
al.\ are not sensitive to emission on scales much larger than 10--30
arcseconds.  Therefore such extended flux is missed.  This is clearly true for
G19.61--0.23A where we have instead used the fluxes presented by Garay et al.\
(1998).  Kurtz et al.\ (1999) demonstrate that such extended emission is
commonplace in compact HII regions.  Such `compact' HII regions tend to show a
compact dense core, which is what was detected by Wood \& Churchwell (1989) and
Kurtz et al.\ (1994), surrounded by low density extended ionised gas emission.
We also do not correct for any dust present in the HII region, which also
acts to depress the derived flux of hydrogen ionising photons.  The work of
Hoare, Roche \& Glencross (1991) suggests this is a reasonable approximation.
Finally we note that we have used the distance estimates given by Wood \&
Churchwell (1989) and Kurtz et al.\ (1994).  There is evidence that these are
on average too large (see, eg, Araya et al.\ 2002, which gives distances for
six of our targets).  However, the effect of a slightly too large distance more
or less compensates for any missing radio emission, so the actual net effect is
small when considering the radio data.

By contrast with the radio data, the observed IRAS fluxes may well 
be too large, since the large IRAS beam can encompass
emission from neighbouring sources given the high source
density in the Galactic plane.  Inspection of higher (18$''$) resolution
21$\mu$m images from the MSX satellite mission (see, eg, Price et al 2001 for
details of this mission) clearly shows the source multiplicity within the IRAS
beam in many cases.  The most discrepant are G5.89--0.39, G19.61--0.23A,
G35.20--1.74 and G45.07+0.13, where the observed MSX 21$\mu$m flux densities
are only 53\%, 65\%, 63\% and 65\% respectively of the IRAS 25$\mu$m flux
density.  In addition, W51d is only a small part of the W51 HII region (and
merges with other sources even in the MSX beam).  Therefore, under the
assumption that most of the bolometric luminosity is reprocessed by dust, the
far infrared luminosity as seen by IRAS gives an upper limit to the true
bolometric luminosity.  In addition, the likely overestimate in the distances
should be borne in mind since this also inflates the estimated IRAS luminosity
(by a factor of $\sim2$ typically if the distances in Araya et al.\ 2002 are
correct).

\section{HII Region Models}
\subsection{Stellar Models}
Any photoionisation model of a HII region requires some assumption as to the
form of the spectral energy distribution of the exciting star(s).  We used the
CoStar model stellar atmospheres as our main input (Schaerer et al.\ 1996a,b).
These have the advantage of being widely available and used in such
photoionisation modelling, and therefore provide a benchmark against which we
can test the observations.  The CoStar models were placed on a physical
luminosity-effective temperature scale by Schaerer \& de Koter (1997) using the
calibrations of Vacca, Garmany \& Shull (1996) who presented the fundamental
relations between observational properties of O stars.  We used the predicted
values for $N_{LyC}$ from Schaerer \& de Koter to normalise our model set.

The Vacca et al.\ calibration may be inaccurate however (see, eg, the
discussion in Herrero 2003, Martins, Schaerer \& Hillier 2002, or Smith, Norris
\& Crowther 2002).  The Vacca et al.\ calibration relied on results from pure
hydrogen and helium non-LTE models and is likely to overestimate the effective
temperature for a given spectral type by as much as 10--15\% (see, eg, Herrero
2003 for an overview). Martins et al.\ (2002) show the likely effect for the
models of most interest to us, the dwarfs, with the effective temperature being
too hot by at least 1500K for all O types.  Crowther et al.\ (2002) and Herrero
(2003) show that the same effect holds for supergiants as well.  There is also
observational evidence that the assumed absolute V band magnitude (and hence
luminosity) derived by Vacca et al.\ is overestimated slightly for the hotter
dwarfs (compare Figure 1 in Smith et al.\ 2002 with Figure 6 in Vacca et al).

In order to estimate the extent of the effect that the use of the CoStar grid
and the Vacca et al.\ calibration has on our results we also used data from a
smaller grid of models presented by Smith et al.\ (2002).  These are based on
the WM-basic models of Pauldrach, Hoffmann \& Lennon (2001), but use the Smith
et al.\ luminosity calibration.  We adopted these models since they span the
range of effective temperature we are interested in as well as being readily
available, though other similar models are now available (see, eg, Herrero 2003
for a summary).  Smith et al.\ (2002) show that the hard ultraviolet flux in
the CoStar models is enhanced compared to the WM-basic models (eg see their
Figure 4).  This is likely due to a fuller treatment of the line blanketing in
the WM-basic models (see also the discussion in Section 4.2 of Giveon et al.\
2002 regarding the comparison of WM-basic models with simple LTE models).

We have also included an extension to the effective temperature sequence for
stars with masses $\ls20$\Msolar.  These are taken from Puxley (1988) and
approximate the behaviour of lower mass stars.  The stellar atmospheres used in
these models are the LTE models from Kurucz (1979).  These models are known to
be incorrect to within a factor of a few.  Therefore we only use these models
to demonstrate why we can largely ignore the lower mass stars in our analysis,
and do not use them in any detailed photoionisation modelling.

\subsection{Observable Properties}
If there is a cluster present, then we need to know the form of the stellar
IMF before we can compare models with observations.  Scalo
(1998) reviews our current understanding of the IMF in depth.  For
our purposes however we adopted the relatively simple IMF proposed by
Salpeter (1955), where the IMF is described solely by a power
law,
\[\psi(M)\propto M^{-2.35}.\]
This is sufficiently close
to our present knowledge of the upper end of the IMF that it
is still a reasonable approximation.  The fact that it is a poor approximation
to the lower end of the IMF is unimportant because the
steepness of the dependence of the ionising photon flux on stellar mass means
we need only consider massive stars in any event.  The number of hydrogen
ionising photons for a given stellar cluster is then
\[  N_{LyC} = \int  n_{LyC}(M) \psi(M) dM \]
where $n_{LyC}$ is the number of hydrogen ionising photons at a given stellar
mass.   A similar relation holds for the total luminosity, $L_{tot}$.

Figure 1 shows how the combinations $n_{LyC}(M) \psi(M)$ and $L(M) \psi(M)$
behave as a function of stellar mass for the CoStar models.  This clearly
demonstrates why we can use the comparison of IRAS and radio data to constrain
the form of the IMF.  The former has a much steeper dependence on $M$ than the
latter, so that $L_{tot}$ is much more sensitive to the presence of lower mass
stars.  Although the details of the functional form differ slightly, the same
is true for the WM-basic models presented in Smith et al.\ (2002) as well.

In Figure 2 we plot the actual comparison of $N_{LyC}$ and $L_{tot}$ for our
sample.  The lower solid curve represents the track followed by a single star
of increasing mass. The points marked are for (a) the dwarf CoStar models
listed in Table 3 of Schaerer \& de Koter (1997) and (b) the hottest 9 dwarf
models listed in Table 1 of Smith et al.\ (2002).  The upper curve presents the
track of a stellar cluster.  The points marked here represent the same spectral
types as for the single stars, but here these are the most massive stars
present in the cluster.  We have normalised these curves so that there is
exactly one star with that spectral type.  Other possible combinations such as
a steeper IMF (as perhaps indicated by the review of Scalo), or a truncated IMF
at both upper and lower mass limits lie somewhere between the two curves shown.
The WM-basic models presented by Smith et al.\ (2002), using their luminosity
calibration, largely overlie the tracks shown for the CoStar models, but with
the points offset approximately 1.5~sub-types towards lower luminosity
than the CoStar models.

We have also plotted the direction in which the observational errors described
in Section 2.2
move the observed luminosity and $N_{LyC}$ points in Figure 2.  The
plotted tracks show the effect of a decrease in the bolometric luminosity by a
factor of two (the extreme suggested by the comparison of the IRAS and MSX data
above), an increase in $N_{LyC}$ by a factor of three (roughly the case
presented by the example of G19.61--0.34A), and a decrease in distance by a
factor of two as suggested by Araya et al.\ (2002).  The distance errors move
the observed data approximately parallel with the model tracks, so any large
discrepancy between the observed data and the models is unlikely to be due to
this source.  Most of the observed data cluster around or above either the
cluster line or single star line in Figure 2.  Errors of the magnitude shown
are certainly sufficient to move all of the sources onto either cluster or
single star tracks.  

We have estimated two characteristic effective temperatures for each of our
sources for both CoStar and WM-basic stellar models.  These are estimated
simply by locating the nearest model that lies directly below the observed data
point.  We have not interpolated these temperatures.  The results are given in
Table 2.  The first value given in Table 2 is that for the single star required
to match the radio data.  The second is the hottest star in the cluster that
provides a best match to the observed radio and far infrared data.  Since none
of the errors move the objects from significantly below the cluster track to
above it, we assume that sources below the cluster track are powered by single
stars only.  For these cases therefore we do not give a type for the most
massive star in a cluster.  The likely observational errors imply that a given
source may actually be up to two sub-types hotter than this and as many as five
sub-types cooler.

Note that although the CoStar and WM-basic model sequences in Figure 2 do not
overlie each other in terms of the actual O sub-type, the same effective
temperatures tend to lie in the same place in these diagrams.  This explains
the rough equivalence between the temperatures of the two model sets given in
Table 2.  Any differences are likely due to the coarseness of the model grids,
especially for the Smith et al.\ models.  In the actual plots of the observed
data we therefore use only the effective temperatures derived from the CoStar
models because the sampling is better.

\subsection{Photoionisation Models}
We used Cloudy version 94.0 (Ferland 2000) to calculate the emission line
properties for our model HII regions.  We ran a grid of models covering the
full range of available stellar effective temperature ($T_{eff}$) for the
CoStar model grid, for all three luminosity classes.  Although we are
interested in young massive stars, there have been suggestions previously that
supergiant models are actually a better match to the observational data.  This
is probably in line with the known properties of some massive young stellar
objects, since there is evidence that their spectra bear a resemblance to
hypergiants (eg Hamann \& Persson 1989).  In addition, we also ran a smaller
set of dwarf and supergiant models using the WM-basic set from Smith et al.\
(2002).

We adopt a `standard' model where we compare the different stellar inputs, with
a flat electron density of $n_e=6000$cm$^{-3}$, a turbulent velocity of
5kms$^{-1}$, inner and outer radii to the nebulae of 0.01~pc and 0.5~pc
respectively (see below) and input stellar models appropriate to solar
metallicity.  Where we vary a given parameter, we kept the other parameters
fixed at the values for this standard unless otherwise noted.  We assume
throughout that the helium abundance is 10\% by number when computing the
models, but allow the metal abundance to vary in line with the stellar model
for the WM-basic model set.

We also varied three other physical parameters in the models we considered in
order to determine which factors were important in influencing the HeI
2$^1$P--2$^1$S/HI Br$\gamma$ ratio.  These were the electron density (and
density structure), the turbulent velocity within the ionised gas and the
metallicity of the exciting star.  We used the the CoStar model set when
considering the effect of density and the microscopic velocity field, and the
WM-basic set when considering metallicity.  The first two parameters are
independent of the nature of the stellar model, whilst only the WM-basic model
set spans sufficient metallicity range to be worth considering.  We constructed
models spanning all five of the metallicities considered by Smith et al.\
(2002).  

We considered a range of turbulent velocities, from 0 to 15kms$^{-1}$.  The
HeI 2$^1$P--2$^1$S line is expected to be sensitive to local velocity
broadening (hence turbulence) because of the resonance pumping from the
2$^1$P--1$^1$S transition (see Paper 1 for more details).  The range chosen
matches the typical range seen in compact HII regions.  For the WM-basic models
we only used a value of 5kms$^{-1}$ throughout.

We considered two possible density models for the HII regions in the CoStar
models: a flat constant density model, with $n_e$ varying between 1000 and
50000cm$^{-3}$, and a model with a flat baseline density $n_e=1000$cm$^{-3}$
with a superposed radial Gaussian density profile, with peak density again
varying up to 50000cm$^{-3}$.  In both cases the inner surface of the cloud was
0.01~pc from the exciting source, and a nominal 0.5~pc outer radius was
adopted. The latter matched the observed radii of our sources to within a
factor of a few in all cases, though the actual model results are relatively
insensitive to this parameter.  We adopted a full width at half maximum of
0.3~pc for the Gaussian component.  The radii were the same in the WM-basic
models, but we only considered the flat density law with $n_e=6000$cm$^{-3}$.

We also considered two sets of models with a larger inner radius.
In the first, we set the inner radius to the cloud
from the exciting star of 0.3~pc, but otherwise used the same set of
parameters as the flat density models.  This radius is actually larger than the
total size of many of the objects being considered, and so this is not a
realistic physical model of these sources.  It can be thought of as a crude way
to approximate the effect of using stellar models with a lower ionisation
parameter since the net effect is to reduce the ionising flux at the HII
region.  We define this parameter as the dimensionless ratio of the photon flux
to the hydrogen density at the inner surface of the HII region (note, not the
Stromgren radius as in the usual definition: our definition agrees with that
used in Cloudy), so that
\[ U = \frac{N_{LyC}}{4\pi\,r^2n_H\,c}\,\,, \]
where $r$ is the inner radius of the HII region, and $n_H$ the total hydrogen
density.  Trials showed that we had to increase this radius to $\sim0.3$~pc
before an effect was seen in the results (models with inner radius of 0.1~pc
are similar to the models with inner radius of 0.01~pc).  This corresponds to a
drop of $\sim1000$ in the ionisation parameter.  The outer radius for these
models was increased to 2~pc.  The second large inner radius set we considered
used CoStar models with an inner radius of 1~pc and an outer radius of 5~pc.
This model assumed a flat electron density profile, and only a model with
$n_e=50000$cm$^{-3}$ and turbulent velocity of 5kms$^{-1}$ was calculated.

\subsection{Model Results}
Figure 3 and 4 show the results from our Cloudy simulations for the different
sets of stellar inputs (luminosity class and model type) for the key HeI
2$^1$P--2$^1$S (Figure 3) and HeI 4$^{3}$D--3$^{3}$P (Figure 4) to HI
Br$\gamma$ line ratios.  In both Figure 3(a) and 4(a) the results are presented
for our standard model for dwarf, giant and supergiant CoStar model
atmospheres, and for dwarf and supergiant WM-basic model atmospheres. The
sampling of the WM-basic models is coarser than for the CoStar models which may
partly explain the shift in the peak between the CoStar and WM-basic dwarf
models.  The CoStar models as a whole show greater variance with type than the
WM-basic models for both line ratios however.  As noted in Section 3.1 this is
probably a reflection of the fact that the CoStar models have insufficient line
blanketing, which is of greater importance in stars with larger mass loss rate
(eg supergiants as opposed to dwarfs).  The net result is a spectrum that is
`harder' at a given effective temperature than it should be, particularly
in the far ultraviolet.  By comparison,
the dwarf and supergiant WM-basic models give very similar predictions
(identical in Figure 4(a), so only the dwarf models are actually plotted).

In Figures 3(b) and 4(b) we consider the effect of having a cluster of stars
rather than a single star.  We again used the standard model and the CoStar
dwarf model atmospheres.  The cluster was constructed so that one star with the
spectral type of the maximum effective temperature was present, and the
normalisation for the other stars was then set by matching the observed
$N_{LyC}$ and $L$ for a cluster as in Figure 2.  There is clearly little
difference in the HeI/HI ratios between the single-star and the cluster case.
This is in line with Figure 1, which clearly shows that the number of helium
ionising photons drops sharply as the stellar mass decreases below 40\Msolar
(equivalent to $T_{eff}\ls42000$K for the CoStar dwarf models).  These results
suggest a prior understanding of the IMF is not required in order to place a
constraint on the hottest star present using these ratios.

In Figure 5 we show the results from our standard model for a different line
ratio to show that large differences do exist between the dwarf and supergiant
Cloudy models and the WM-basic models.  The line ratio plotted is that of
15.6$\mu$m [NeIII] with 12.8$\mu$m [NeII].  This ratio is commonly obtained
from spectroscopy obtained by the ISO satellite of compact HII regions (eg
Morisset et al.\ 2002, Giveon et al.\ 2002).  These model results clearly show
much wider variation in behaviour than any of the HeI/HI line ratios.  We have
also shown how metallicity can effect such line ratios as well by examining the
variation with metallicity for the WM-basic dwarf models.  The result is
largely as expected if higher metallicity leads to more line blanketing in the
far ultraviolet, and hence a softening of the spectral energy distribution (the
difference occurs at energies larger than 2.5 Rydberg, due to the change in the
species causing the line blanketing: see the extended discussion of this issue
in Giveon et al.\ 2002 for fuller details).  These results highlight both a
strength and weakness of the HeI/HI ratios in placing constraints on the
effective temperatures of the stars present.  The main strength is that the
results you would obtain are largely independent of the stellar model chosen
and metallicity (since the helium fraction alters with metallicity by a much
smaller amount than any of the metals).  The main weakness is that it can place
only weak constraints on the spectral energy distribution of the stellar
atmosphere itself.

Figure 6 shows how the metallicity of the underlying star effects the results.
Again we show both the HeI 2$^1$P--2$^1$S (a) and HeI 4$^{3}$D--3$^{3}$P (b) to
HI Br$\gamma$ line ratios.  There is a weak metallicity dependence in the
former, probably as a result of the increase of the neutral helium fraction in
the nebula due to the softening of the ultraviolet flux with increasing
metallicity.  The resonance of the 2$^1$P--1$^1$S transition is crucially
dependent on this neutral helium fraction as first noted by Shields (1993).  By
contrast the changes are relatively small in the pure recombination line ratio.

Figure 7 shows how the ratio varies as a function of the turbulent velocity.
Increasing the turbulent velocity depresses the ratio, as
was seen for the models for the PN presented in Paper 1.  This is simply due to
the fact that the resonance scattering of the 2$^1$P--1$^1$S HeI line is more
likely to lead to capture by dust or atomic hydrogen if the line is much
broader than the thermal value.  The differences seen are of the same order of
magnitude here as in Paper 1, as expected given that this is a test of
the physics in Cloudy and not really of the form of the input stellar models.

Figure 8 shows the effect on the HeI 2$^1$P--2$^1$S/HI Br$\gamma$ ratio of
changing the electron density (the pure recombination line ratio is independent
of density).  Figure 8(a) shows the data for a flat density distribution, and
Figure 8(b) the data for the Gaussian density distribution.  There is clearly
little difference between the two, and little apparent dependence on density.
This is perhaps surprising since it is well known that the 2$^1$S level of
helium can be pumped by collisional excitation from the triplet 2$^3$S
`ground-state' (see, eg, Doherty et al.\ 1994).  In practice these Cloudy
models predict the 2$^3$S level is primarily depopulated by photoionisation
instead (see also the discussion in Clegg \& Harrington 1989) however.

Figure 9 shows the results for a flat model with a larger inner nebular radius.
The solid lines show the result of increasing the radius to 0.3~pc, 
with the same densities as those shown in Figure 8(a),
whilst the
dashed line shows the result for an inner radius of 1~pc for a single electron
density.  It must be stressed that these models should only be appropriate for
older more extended HII regions.  However, it is clear that a reduction in the
ionisation parameter at the inner edge of the nebula
does have a striking effect on the line ratio.  The most
obvious feature is the enhanced density dependence of the HeI
2$^1$P--2$^1$S/HI Br$\gamma$ ratio, clearly indicating the growing importance
of collisional excitation of the 2$^1$S level.  This is a reflection of the
much higher ultraviolet photon density at the inner surface of the HII region
in our other models.  The effect seen in the models with a larger inner radius
should in principle be seen in our `normal' models if we increased the density
significantly beyond 10$^5$cm$^{-3}$ since the collisional rate may then
dominate the photoionisation rate.  Such densities are not in agreement with
the published estimates from radio data however.  It is also worth noting the
fact that the very extended HII region model shows a much shallower decline of
the HeI 2$^1$P--2$^1$S/HI Br$\gamma$ ratio with increasing effective
temperature.

The overall trends seen in the models can be summarised fairly simply.  There
are weak variations in the HeI 2$^1$P--2$^1$S/HI Br$\gamma$ ratio with almost
all the parameters considered, but only the increase of the inner radius
combined with changes in the electron density have a significant impact.  By
contrast, the pure recombination line ratios are largely independent of most of
the factors we have considered.

\section{Observational results}
\subsection{General Trends}
Figure 10 shows the observed HeI 2$^1$P--2$^1$S/HI Br$\gamma$ ratio together
with a representative sample of the CoStar models from Section 3.  We have not
plotted the WM-basic model here since it is very close in appearance to the
CoStar dwarf model (Figure 3).  The behaviour seen is almost completely at
variance with the models.  One of the two lowest temperature sources plotted
here (G5.97--1.17) lie well above the predicted curve.  The nature of this
source is not entirely clear, since it lies very near the exciting star of M8,
Herschel 36.  Stecklum et al.\ (1998) have speculated that the emission
actually arises in material near a young, but much less massive, star which is
being photo-evaporated by the radiation of Herschel 36.  If true, the actual
effective temperature of the exciting star is that of Herschel 36 (an O7V
star), and the point should be shifted to higher $T_{eff}$ on the figure.  The
other key point to note from Figure 10 is that the observed ratio at
$T_{eff}>40000$K is always higher than the theoretical prediction unless we
adopt the lowest ionisation parameter model.  There is also no downturn in the
observed ratio as $T_{eff}$ increases beyond 40000K, as predicted in all the
models.  It must be noted however that the largest inner radius model declines
more slowly than the others and is hence a better, though still poor, match to
the data.

The `pure' recombination line ratios are shown in Figure 11.  We have also
shown a representative spread of the stellar models in Figures 11(a) and (b),
with dwarf CoStar models in (a) and dwarf WM-basic models in
(b).  Since the 7$^{3,1}$G--4$^{3,1}$F/4$^{3}$D--3$^{3}$P ratio is effectively
constant, the WM-basic model for (a) would look the same as in (b) but with the
plateau value the same as for the CoStar models (and vice-versa for the CoStar
models not shown in (b)).  The low ionisation parameter/large inner radius
models shown in Figure 10 are not shown in Figure 11 since they given
essentially the same results as the dwarf models plotted.

In general, these models agree reasonably well with the dwarf models at
$T_{eff}>38000$K.  There is some discrepancy between dwarf models and
observation at lower effective temperature.  Again, in part, this is due to
G5.97--1.17.  There is some evidence from the data however that the `plateau'
in these ratios arrives at a slightly cooler effective temperature than the
models predict, though the difference is certainly within the errors on the
adopted object effective temperatures given in Table 2.  The other truly
discrepant
point is that for G5.89--0.39, which lies well below the predicted model curve
for the HeI 7$^{3,1}$G--4$^{3,1}$F/HI Br$\gamma$ ratio.  It may also lie below
the model HeI 4$^{3}$D--3$^{3}$P/HI Br$\gamma$ ratio but the observational
error for this ratio is too large to draw any conclusions.  We discuss this
source in more detail below.  Overall, however, our results are in accord with
the study of a large sample of optically visible, and hence more evolved HII
regions, by Kennicutt et al.\ (2000).  They found that the ratio of optical HeI
and HI lines were in general agreement with the predictions from Cloudy using
main sequence models for the exciting stars.  Our results tend to indicate that
the same is true regardless of the stellar model used.

It is worth briefly considering whether the data might match one of the CoStar
supergiant models shown in Figures 3 and 4.  There have been suggestions that
such models sometimes give a better match between theory and observations to
other line ratios in compact HII regions (see Section 4.2 below for an
example).  Note from Figures 3 and 4 there is no difference between dwarf and
supergiant WM-basic models, so we do not consider these here.  Of course the
observational data as plotted in Figure 10 and 11 assume the exciting stars
have effective temperatures compatible with the CoStar dwarf models.  In
practice, if we define the spectral type using CoStar supergiant models in the
same fashion as Figure 2, then all of the sources fitted by a single
exciting star actually lie well below the end of the CoStar model sequence.
The others have types ranging from O7I/O8.5I downwards depending on whether a
single star model or a cluster is used.  Therefore the observational data
should all be shifted to the left by $\sim10000$K when comparing the
observational data in Figures 10 and 11 with the supergiant models in Figures 3
and 4.  This clearly makes most of the observational data disagree completely
with the model.  The problem is actually exacerbated if we consider a cluster
of exciting stars.  It seems clear that supergiant CoStar models are not a
sensible match to the HeI/HI line ratio data in general.

Figure 12 shows the ratio of the HeI 2$^1$P--2$^1$S line with the two `pure'
HeI recombination lines.  We showed in Paper 2 that this ratio helped remove
any abundance effects when comparing observational data with the predictions
from Cloudy assuming a single helium abundance.  The data here however are
actually more consistent with a flat line than the predicted curves.  The
results therefore show that the cause of the high HeI 2$^1$P--2$^1$S/HI
Br$\gamma$ ratio is not an abundance effect, but is intrinsic to the strength
of the HeI 2$^1$P--2$^1$S line.  Even the model with a 1~pc inner radius and
high electron density fails to explain the observed data.  These results are
completely different from those for the planetary nebulae, where these HeI/HeI
ratios generally showed reasonable agreement between model and observation.

We must stress however that these HII regions have electron densities that are
generally a factor of $\sim$10 higher than in the planetary nebulae.  It is
worth noting that there is a small discrepancy between the observed and model
HeI 2$^1$P--2$^1$S/HI Br$\gamma$ line ratio even in the planetary nebulae (see
Figure 4 and 5 of Paper 2), in the sense that the HeI 2$^1$P--2$^1$S line is
stronger than expected by 10--20\% from the mean model trend.  As noted above,
we expect the HeI 2$^1$P--2$^1$S to be stronger 
if the collisional excitation rate from 2$^3$S exceeds
the photoionisation rate.  If the Cloudy estimation of these two rates is
inaccurate (or equally if our assumed model inputs for inner cloud radius and
electron density are too small), then it is possible that the collisional
excitation rate actually dominates in these compact HII regions in reality.
However, even allowing for this, it is notable that the HeI 2$^1$P--2$^1$S/HI
Br$\gamma$ line ratio does still decrease at higher effective temperatures for
the planetary nebulae, and we do not see that effect here.  It seems clear that
at least one other factor must play a role in explaining the poor match between
model and observed HeI 2$^1$P--2$^1$S/HI Br$\gamma$ line ratios for the compact
HII regions.

The most important caveat in these general conclusions is that we rely on the
data shown in Figure 2 to derive $T_{eff}$.  We have used the data as presented
in Wood \& Churchwell (1989) and Kurtz et al.\ (1994).  As noted in Section 2,
there are reasons for believing that at least some of the distance estimates
they use are too large.  A lower distance estimate may help to explain some of
the problem for those sources in Figure 10 which lie at $T_{eff}>40000$K, but
which lie well above the model HeI 2$^1$P--2$^1$S/HI Br$\gamma$ ratio, since it
would move the observed points to lower $T_{eff}$.  However, it cannot be too
much lower, since Figure 11 shows reasonable agreement between model and data.

\subsection{Individual Sources: G29.96--0.02}
There are several parameters that we have not discussed so far that may affect
our conclusions.  These are best illustrated by considering specific examples.
A useful example is G29.96--0.02.  This object has been discussed at length in
the literature.  There are good arguments for believing that it may be at a
distance of 6~kpc rather than the 9~kpc given in Wood \& Churchwell (1989), or
even the 7.4~kpc estimated by Churchwell, Walmsley \& Cesaroni (1990) (see
Pratap, Megeath \& Bergin 1999 for a fuller discussion regarding the distance
to this object).  G29.96--0.02 is more extended in the radio than indicated by
the radio map of Wood \& Churchwell (1989).  Fey et al.\ (1995) derive a value
of $N_{LyC}$ which is $\sim$50\% higher than Wood \& Churchwell, largely
compensating for any reduction in distance with regard to this indicator (see
also the discussion in Section 2.2).  Using the Fey et al.\ data and a distance
of 6~kpc $\log N_{LyC}=49.1$s$^{-1}$.  The bolometric luminosity however does
decrease to $\log L = 5.95\Lsolar$.  The net results is a drop of half a
sub-type in the best fitting stellar dwarf models, with a cluster still being
preferred over a single star (see Figure 2).  The expected pure HeI/HI
recombination line ratio for such a model is still on the plateau seen at high
$T_{eff}$ in Figure 11, in agreement with observation.

There have been many previous indirect estimates of the spectral type of the
exciting star(s) in this HII region, in the same spirit as this paper but
primarily using mid-infrared forbidden line data.  The most recent of these by
Morisset et al.\ (2002), based on ISO data, concluded that the best matched
exciting star was actually a relatively cool supergiant ($\sim$O9.5I, or
$T_{eff}\sim30000$K using CoStar models).  This actually disagrees with our
pure recombination line ratios (Figure 11, and see the discussion in Section
4.1), and hence we can rule that model out.  Such a model can also probably be
ruled out on the basis of other evidence.  The presence of young hot cores
(eg.\ Cesaroni et al.\ 1998, De Buizer et al.\ 2002) near the HII region
strongly suggest that G29.96--0.02 is part of a complex of star formation
taking place on a larger scale (also consistent with the complexity seen on
larger scales at radio wavelengths by Kim \& Koo 2001).  Furthermore our own
work on the kinematics of this object (Lumsden and Hoare 1996, 1999) indicate
an evolved star travelling through a molecular cloud is not a good match to the
data.

It is worth briefly considering why Morisset et al.\ reached the conclusion
they did.  There are several line ratios in the mid and far-infrared that are
abundance independent.  These include, limiting ourselves to the better studied
mid-infrared ratios only, 6.99$\mu$m [ArII]/8.99$\mu$m [ArIII], 18.7$\mu$m
[SIII]/10.5$\mu$m [SIV], and 12.8$\mu$m [NeII]/15.6$\mu$m [NeIII].  The latter
two are of particular interest since S$^{++}$ and Ne$^+$ have approximately the
same ionisation potential as He$^0$.  Therefore, in principle, these ratios add
extra constraints on the form of the stellar spectral energy distribution
beyond 504\AA.  Here we will only consider the 12.8$\mu$m [NeII]/15.6$\mu$m
[NeIII] ratio in detail.  Morisset et al.\ give an extinction corrected
observed ratio of 3.75$\pm0.3$ (the observed ratio is close to 2).  As can be
seen from the inspection of Figures 4 and 5 a ratio this low, together with
the observed HeI/HI ratio actually rules out all CoStar models.  Morisset et
al.\ adopted a low value for the HeI/HI ratio, based on weak helium radio
recombination line data from Kim and Koo (2001), allowing a fit to the data
with cool CoStar supergiant models.  Our result is certainly more reliable, and
hence we can exclude this model as previously noted.

Indeed, it is not just G29.96--0.02 where the combination of our near infrared
data and published mid-infrared data leads to problems for the CoStar models.
If we accept the best matching CoStar models to the mid-infrared data all of
G5.89--0.39, G29.96--0.02, G43.89--0.78, G45.07+0.13, G45.12+0.13, G45.45+0.06
and W51d would have temperatures inconsistent with the HeI/HI ratios presented
in Figure 11 (see the ISO data presented in Giveon et al., or the IRAS LRS data
in Simpson \& Rubin 1990).  There is a caveat here, since all of these data
were acquired with a large beam and sample most if not all of any extended
emission present, whereas our results are largely sensitive only to the dense
cores of the HII regions observed.  However, this overall trend does indicate a
problem with the CoStar models, in the sense that the far ultraviolet spectral
energy distribution is harder than observed.

By comparison, the WM-basic models allow a solution to both near and
mid-infrared line ratios for $T_{eff}\sim35000-37000$K (the exact temperature
is somewhat metallicity dependent).  This range gives reasonable agreement
between model and data for all three of the mid-infrared line ratios listed
above for the case of G29.96--0.02.  This is perhaps surprising considering we
have not run a detailed physical model such as the one Morisset et al.\ use.
However, the agreement is basically a reflection of the fact that at a given
effective temperature the WM-basic models are softer in the far ultraviolet
region of the spectrum than the CoStar models (see also Section 4.2 of Giveon
et al.\ 2002), and appear to be a better match to the data for most HII regions
(again see Giveon et al).

The most useful direct constraint for G29.96--0.02 lies in a spectrum of the
main exciting star (Watson \& Hanson 1997, Hanson 1998, Kaper et al.\ 2002).
The published spectral type for the star varies significantly between these
references (as hot as O3 in Kaper et al., as cool as O8 in Watson \& Hanson),
but the most likely range is O4--O6 from the strengths of the CIV, HeI and HeII
lines present (Hanson, private communication).  This range is consistent with
the photometry of Watson et al.\ (1997), especially if we use a lower value for
the extinction at $K$ of 1.6 (as proposed by Morisset et al.\ 2002, and as
determined from a preliminary calibration of our own data for studying the
extinction law).  Although this would have an absolute magnitude at the limit
for a main sequence O4 star (eg.\ Vacca et al.\ 1996 or Hanson, Howarth \&
Conti 1997), it is still just consistent with that type if the distance is only
6kpc (the cooler types are all consistent).  It is worth noting that this range
of spectral type is still hotter than expected from the photoionisation models.
This is not surprising given the uncertainty over the effective
temperature/spectral type relation however (and the implication from all of the
more recent work that this scale needs to shift to cooler temperatures).  It
does however emphasise that we should still treat the existing stellar models
with some caution.

\subsection{Individual Sources: G5.89--0.39}
Another useful individual example to consider is G5.89--0.39.  The distance to
this source may be slightly less than reported in Wood \& Churchwell (1989) but
the effect is small, and the accuracy of the distance measure for this
particular source is good (see Acord, Churchwell \& Wood 1998 for a fuller
discussion of the distance).  For either distance however, there is a deficit
of helium ionising photons in this source, as is clear from Figure 11(a), where
the observed HeI 7$^{3,1}$G--4$^{3,1}$F/HI Br$\gamma$ ratio is well below both
any model prediction and the observed values for other sources with the same
expected effective temperature.  This discrepancy cannot be explained by errors
in the assignment of a spectral type, since the distance is well known, and the
radio emission compact.  It is perhaps noteworthy that G5.89--0.39 may be the
youngest source in this sample.  The HII region
has an age of only a 600 years from
estimates of its rate of expansion (Acord et al.\ 1998).  There is also a
substantial molecular outflow (Acord, Walmsley \& Churchwell 1997) and an
infrared reflection nebulosity associated with it (Lumsden \& Hoare in
preparation), symptomatic of a rather young source.

For the case of G5.89--0.39 it is tempting to consider the possibility that the
stellar models used are actually inappropriate, and its young age may be the
clue here.  As noted previously, there is some evidence that massive young
stellar objects appear to share spectral characteristics with hypergiants
(Hamann \& Persson 1989, Bunn et al.\ 1995).  Both classes of objects have
dense slow moving stellar winds, giving rise to this similarity.  It is also
true that most massive young stellar objects do not show significant HII
regions, although they are luminous enough to generate them.  It is assumed
that the winds act as sinks for much of the ultraviolet flux from the star,
suppressing the ability to form an HII region, and probably also softening the
spectral energy distribution over that seen in main sequence stars.  If the
same behaviour was also shown by the central stars of very young HII regions
(perhaps because the atmosphere of the star has not fully attained its main
sequence structure) then it might explain a case such as this.  G5.89--0.39 is
actually more luminous than the stars studied by Bunn et al., so
even some leakage of ultraviolet photons could lead to a substantial HII
region.  Of course, we cannot prove this speculation from our current data.

\section{Conclusions}
We have presented models for the infrared helium and hydrogen recombination
line ratios in compact HII regions.  We fail to find any model that can
accurately reproduce the HeI 2$^1$P--2$^1$S line strength as a function of
effective temperature.  We therefore would discourage anyone from using this
ratio as a constraint on the IMF, and any evidence derived from the literature
about the IMF in a particular system that relies on this ratio should largely
be discounted.

Our results do show reasonable agreement between model and observation for pure
recombination lines such as HeI 4$^{3}$D--3$^{3}$P or 7$^{3,1}$G--4$^{3,1}$F
when ratioed with HI Br$\gamma$ as long as we use dwarf CoStar or WM-basic
models.  This indicates that these are in principle a useful measure of the
stellar effective temperature of the hottest star present.  The main drawback
with these ratios is the relatively small range of effective temperature over
which they are useful (approximately over the range B0 to O8 for main sequence
CoStar models for example).  However, at least some of the galaxies in which
the 1.7007$\mu$m HeI 4$^{3}$D--3$^{3}$P line has been detected do have values
of the ratio within this range (see, eg, Table 4 of Vanzi \& Rieke 1997).  It
should certainly help to address the issue of whether there is evidence for an
anomalous IMF from infrared data (see, eg, the review by Scalo 1998).

We have outlined possible reasons for the discrepancies seen between the
observations and the models for the HeI 2$^1$P--2$^1$S/HI Br$\gamma$ ratio.
The first, and perhaps most likely, is that the relative importance of
photoionisation and collisional excitation from the $2^3$S HeI state may be
incorrectly estimated by our HII region models.  Clearly, from Figure 9, models
in which the collisional depopulation of $2^3$S is dominant come closer to
matching the observed data.  However, it is still clear from Figure 12 that the
HeI/HeI ratios are closer to a flat function of effective temperature than the
declining function predicted by Cloudy (in agreement with the observations of
the planetary nebulae in Papers 1 and 2).  It seems likely that errors in the
2$^3$S de-population rate may give rise to part but not all of the discrepancy
seen.

Perhaps, instead, the HeI 2$^1$P--2$^1$S line is telling us something about the
nature of the ultraviolet spectral energy distribution of the exciting star?
The current uncertainty in the model stellar atmospheres for massive stars (eg
Martins et al.\ 2002) is unlikely to resolve this problem entirely however,
since we have shown that the HeI/HI line ratios have a relatively weak
dependence on model when comparing CoStar and WM-basic dwarf models.  The
effective reduction in luminosity given by any re-calibration of the spectral
type/effective temperature relation to cooler temperatures may help to lessen
the differences between the model HeI 2$^1$P--2$^1$S/HI Br$\gamma$ ratio and
the observed data however.  

Models with a more realistic treatment of the stellar wind (such as the
WM-basic set) also help to reduce the discrepancy between the pure HeI/HI
recombination ratios and the mid-infrared forbidden line ratios, since the
overall shape of the far ultraviolet continuum is clearly crucial in that
regard.  They also help in closing the gap between the direct spectral type
seen in G29.96--0.02 and the inferred type from the photoionisation models.
Similarly, consideration of changes in metallicity of the stellar model do not
effect the HeI/HI ratios significantly, but can effect forbidden line ratios
(eg Figure 5).  There is evidence for a gradient of metallicity with
Galactocentric radius, with metallicity increasing inwards and excitation
correspondingly decreasing (eg Giveon et al.\ 2002).  Many of our sources lie
within the solar circle and are likely to have exciting stars with higher than
solar metallicity.

The most discrepant data point for all three helium lines is probably the
youngest of the HII regions in the sample used here.  We have outlined the
possibility for this case that the exciting star may actually resemble those
massive young stellar objects studied by Bunn et al.\ (1995).  In this case a
considerable fraction of the emitted ultraviolet flux is actually reprocessed
into non-ionising radiation by a dense stellar wind reducing the excitation in
the nebular gas.  For the moment this must remain purely speculative however.

The easiest way of testing whether our results indicate some residual problems
within the photoionisation model that are not otherwise apparent from Papers 1
and 2 or a deficiency on the assumed stellar models would be to obtain $K$ band
spectra of older, more evolved HII regions, such as those studied by Kennicutt
et al.\ (2000).  More evolved regions are less dense, and larger, and hence any
residual uncertainty in the depopulation route from 2$^3$S should be removed.
If the same effect as seen here were repeated then it would tend to indicate
that even the latest stellar models are still not well matched to the actual
ultraviolet continuum of OB stars.

\section{Acknowledgments}

SLL acknowledges the support of PPARC through the award of an Advanced Research
Fellowship.  We would like to thank the referee Paul Crowther for his helpful
comments, Margaret Hanson for her comments on the likely spectra type of the
exciting star in G29.96--0.02 and Richard Norris for making available his code
to read the WM-basic models into Cloudy.

\parindent=0pt

\vspace*{3mm}

\section*{References}
\begin{refs}
\mnref{Acord, J.M., Walmsley, C.M., Churchwell, E., 1997, \apj, 475, 693}
\mnref{Acord, J.M., Churchwell, E., Wood, D.O.S.,  1998,  \apj,  
495,  L107}
\mnref{Araya, E., Hofner, P., Churchwell, E., Kurtz, S., 2002, \apjs, 138,
	63}
\mnref{Bunn, J.C., Hoare, M.G., Drew, J.E.,  1995,  \mnras,  272, 346}
\mnref{Cesaroni, R., Hofner, P., Walmsley, C.M., Churchwell, E., 1998, 
	\aaa, 331, 709}
\mnref{Churchwell, E., Walmsley, C.M., Cesaroni, R., 1990, \aas, 83, 119}
\mnref{Clegg, R.E.S., Harrington, J.P.,  1989,  \mnras,  239,  869}
\mnref{Crowther, P.A., Hillier, D.J., Evans, C.J., Fullerton, A.W.,
	De Marco, O., Willis, A.J., \apj, 2002, 579, 774}
\mnref{De Buizer J.M., Watson A.M., Radomski J.T., Pina R.K., Telesco C.M., 
	2002, \apj, 564, L101}
\mnref{Doherty, R.M., Puxley, P., Doyon, R., Brand, P.W.J.L.,  
1994,  \mnras,  266,  497}
\mnref{Doyon, R., Puxley, P.J., Joseph, R.D.,  1992,  \apj,  397,  117}
\mnref{Ferland, G.J., 1999,  \apj,  512,  247}
\mnref{Ferland, G.J., 2000, Hazy, a brief introduction to Cloudy 94.00a,
University of Kentucky Department of Physics and Astronomy 
        Internal Report.}
\mnref{Fey, A.L., Gaume, R.A., Claussen, M.J., Vrba, F.J.,  1995,  
\apj,  453,  308}
\mnref{Garay, G., Moran, J.M., Rodriguez, L.F., Reid, M.J.,  1998,  
	\apj,  492,  635}
\mnref{Giveon, U., Sternberg, A., Lutz, D., Feuchtgruber, H., Pauldrach, 
A.W.A., 2002, \apj,  566, 880} 
\mnref{Hamann, F., Persson, S.E.,  1989,  \apjs,  71,  931}
\mnref{Hanson, M.M., 1998, in Boulder-Munich II: Properties of 
	Hot, Luminous Stars, ed. Howarth, I., ASP Conference 
	Series vol. 131, 1998, p. 1}
\mnref{Hanson, M.M., Howarth, I.D., Conti, P.S.,  1997,  \apj,  489, 
 698}
\mnref{Herrero, A., 2003, A Massive Star Odyssey, from Main Sequence to
	Supernova, Proceedings IAU Symposium 212, eds van der Hucht, K.,
	Herrero, A., Esteban, C.}
\mnref{Hoare, M.G., Roche, P.F., Glencross, W.M.,  1991,  \mnras,  
	251,  584}
\mnref{Kaper, L., Bik, A., Hanson, M.M., Comeron, F., 2002, 
 	ASP Conf. Ser. 267, Hot Star Workshop III: The Earliest Stages
	of Massive Star Birth, ed Crowther, P.A., p 95}
\mnref{Kennicutt, R.C., Bresolin, F., French, H., Martin, P.,  2000,  
\apj,  537,  589}
\mnref{Kim, K., Koo, B.,  2001,  \apj,  549,  979}
\mnref{Kurtz, S., Churchwell, E., Wood, D.O.S.,  1994,  \apjs,  91,  
659}
\mnref{Kurtz, S.E., Watson, A.M., Hofner, P., Otte, B.,  1999,  \apj, 
 514,  232}
\mnref{Kurucz, R.L., 1979, \apjs, 40, 1}
\mnref{Landini, M., Natta, A., Salinari, P., Oliva, E., 
	Moorwood, A.F.M., 1984, \aaa, 134, 284}
\mnref{Lumsden, S.L., Hoare, M.G.,  1996,  \apj,  464,  272}
\mnref{Lumsden, S.L., Hoare, M.G.,  1999,  \mnras,  305,  701}
\mnref{Lumsden, S.L., Puxley, P.J., 1996, \mnras, 281, 493}
\mnref{Lumsden, S.L., Puxley, P.J., Hoare, M.G.,  2001a,  \mnras,  
320,  83}
\mnref{Lumsden, S.L., Puxley, P.J., Hoare, M.G.,  2001b,  \mnras,  
328,  419}
\mnref{Martins, F., Schaerer, D., Hillier, D.J.,  2002,  \aap,  382,  
999}
\mnref{Morisset, C., Schaerer, D., Martin-Hernandez, N.L., Peeters, E., Damour,
	F., Baluteau, J.-P., Cox, P., Roelfsema, P., 2002, \aaa, 386, 558}
\mnref{Pauldrach, A.W.A., Hoffmann, T.L., Lennon, M., 2001, \aaa, 375, 161}
\mnref{Pratap, P., Megeath, S.T., Bergin, E.A., 1999, \apj, 517, 799}
\mnref{Price, S.D., Egan, M.P., Carey, S.J., Mizuno, D.R., Kuchar, T.A.,
         2001, AJ, 121, 2819}
\mnref{Puxley, P.J., 1998, Ph.D. thesis, Univ. of Edinburgh}
\mnref{Salpeter, E.E.,  1955,  \apj,  121,  161}
\mnref{Scalo, J.,  1998,  ASP Conf. Ser. 142: The Stellar Initial Mass 
	Function (38th Herstmonceux Conference),   201}
\mnref{Schaerer, D., de Koter, A., Schmutz, W., Maeder, A.,  1996a,  
\aap,  310,  837}
\mnref{Schaerer, D., de Koter, A., Schmutz, W., Maeder, A.,  1996b,  
\aap,  312,  475}
\mnref{Schaerer, D., de Koter, A.,  1997,  \aap,  322,  598}
\mnref{Shields, J.C.,  1993,  \apj,  419,  181}
\mnref{Simpson, J.P., Rubin, R.H.,  1990,  \apj,  354,  165}
\mnref{Smith, L.J., Norris, R.P.F., Crowther, P.A., 2002, \mn, in press}
\mnref{Stecklum, B., Henning, T., Feldt, M., Hayward, T.L., Hoare, M.G., 
	Hofner, P., Richter, S., 1998, \aj,  115, 767} 
\mnref{Thompson, R.I., Tokunaga, A.T.,  1980,  \apj,  235,  889}
\mnref{Vacca, W.D., Garmany, C.D., Shull, J.M.,  1996,  \apj,  460,  
914}
\mnref{Vanzi, L., Rieke, G.H.,  1997,  \apj,  479,  694}
\mnref{Watson, A.M., Hanson, M.M., 1997, \apj, 490, 165}
\mnref{Wood, D.O.S., Churchwell, E.,  1989,  \apjs,  69,  831}
\end{refs}

\onecolumn

\noindent{\bf Table 1:}
The observed HeI/HI line ratios after correction for extinction and atmospheric
absorption.  Ratios with HI Br$\gamma$ are shown.  There are no
suitable observational data for entries marked --.  Note that Doherty et al.\
(1994) do not quote errors on $\tau_{{\rm Br}\gamma}$.  The correction factors
for the 2.058$\mu$m line represent the additional correction necessary
to recover the `true' 2.058$\mu$m line flux.  The quoted ratios should
be divided by this factor to derive the observed ratios.

\begin{tabular}{lccccc}
Name & \multicolumn{3}{c}{HeI/HI Br$\gamma$ ratio} & $\tau_{{\rm Br}\gamma}$ &
   2.058$\mu$m correction\\
   & 2$^1$P--2$^1$S & 7$^{3,1}$G--4$^{3,1}$F & 4$^{3}$D--3$^{3}$P & factor\\
G5.89--0.39  & 0.757$\pm$0.017 & 0.024$\pm$0.005 & 0.092$\pm$0.051 
& 3.80$\pm$0.18&1.06\\
G5.97--1.17  & 0.688$\pm$0.032 & 0.030$\pm$0.008 &   --   & 1.72        &1.28 \\
G19.61--0.23A&        --       & 0.045$\pm$0.005 \\
G28.29--0.36 &        --       & 0.008$\pm$0.003 & --\\
G29.96--0.02 & 0.938$\pm$0.007 & 0.040$\pm$0.007 & 0.103$\pm$0.009
& 1.86$\pm$0.07 &1.09\\
G35.20--1.74 & 1.046$\pm$0.005 & 0.039$\pm$0.003 & 0.101$\pm$0.017 
& 2.50$\pm$0.03&0.95\\
G43.89--0.78 & 0.788$\pm$0.052 & 0.041$\pm$0.004 & -- & 2.45&1.11\\
G45.07+0.13  & 0.820$\pm$0.027 & 0.046$\pm$0.010 &  --            &
4.83$\pm$0.16 &0.99\\
G45.12+0.13  & 1.166$\pm$0.003 & 0.037$\pm$0.006 & 0.109$\pm$0.007 
& 1.19$\pm$0.01&1.05\\
G45.45+0.06  & 0.790$\pm$0.050   & 0.046$\pm$0.007 & 0.083$\pm$0.030 & 2.31$\pm0.54$&0.96\\
K3-50a       & 0.911$\pm$0.024 & --              & 0.104$\pm$0.020 
& 1.53$\pm$0.01&1.17\\
W51d         & 1.162$\pm$0.011 & 0.046$\pm$0.001 & 0.109$\pm$0.013 
& 1.93$\pm$0.07&1.08\\

\end{tabular}

\vspace*{10mm}

\noindent{\bf Table 2:} Properties of our observational sample.  
The observed far infrared luminosities and ionising photon rates (from the
radio) for our sample are taken from Wood \& Churchwell (1989) and Kurtz,
Churchwell \& Wood (1994), except for G19.61--0.23A where we used the 22GHz
flux from Garay et al.\ (1998).  The stellar effective temperatures are taken
from Figure 2 and represent the best matching CoStar and WM-basic models.  The
cluster temperature represents the temperature of the hottest star present.
Where the cluster temperature is marked -- the observed data are actually a
better match to a single star model (ie they lie significantly below the
cluster line), and we do not consider a cluster model for that source.  The
`best' matching model is taken simply by dropping a vertical line down from the
source onto the cluster and single star lines in Figure 2.  Note we have
chosen the nearest model in the set and not interpolated between models in
deriving these values.  

\begin{tabular}{lrccccc}
Name  & $L$ ($10^4$\Lsolar) & $N_{LyC}$ (log$_{10}\gamma$/s) &
  \multicolumn{2}{c}{CoStar $T_{eff}$ (K)} & 
\multicolumn{2}{c}{WM-basic $T_{eff}$ (K)} \\
  &     &   & \multicolumn{1}{c}{Single star} & \multicolumn{1}{c}{Cluster}
& \multicolumn{1}{c}{Single star} & \multicolumn{1}{c}{Cluster}\\
G5.89--0.39  &   30.0 & 48.65 & 37170 & -- & 37200 & --\\
G5.97--1.17  &    8.4 & 47.40 & 32060 & -- & 28500 & --\\
G19.61--0.23A & 26.1 & 48.31 & 35900 & -- & 34600 & --\\
G28.29--0.36 & 11.7 & 47.81 & 32060 & -- & 32300 & --\\
G29.96--0.02 &198.0 & 49.34 & 43560 & 39730 & 45700 & 40000\\
G35.20--1.74 & 28.3 & 48.43 & 35900 & -- & 34600 & --\\
G43.89--0.78 & 27.6 & 48.74 & 38450 & -- & 37200 & --\\
G45.07+0.13  &142.4 & 48.75 & 38450 & 35900 & 37200 & 34600 \\
G45.12+0.13  &166.0 & 49.53 & 46120 & 41010 & 50000 & 40000 \\
G45.45+0.06  &144.0 & 48.54 & 35900 & 34620 & 34600 & 32300 \\
K3-50a       &228.0 & 49.29 & 43560 & 39730 & 45700 & 40000 \\
W51d         &341.0 & 49.42 & 44840 & 39730 & 45700 & 40000 \\

\end{tabular}

\newpage

\vspace*{-1cm}

\begin{center}
\begin{minipage}{6.5in}{
\psfig{file=Figure1.ps,width=6.5in,angle=0,clip=}}\end{minipage}

\begin{minipage}{\textwidth}{
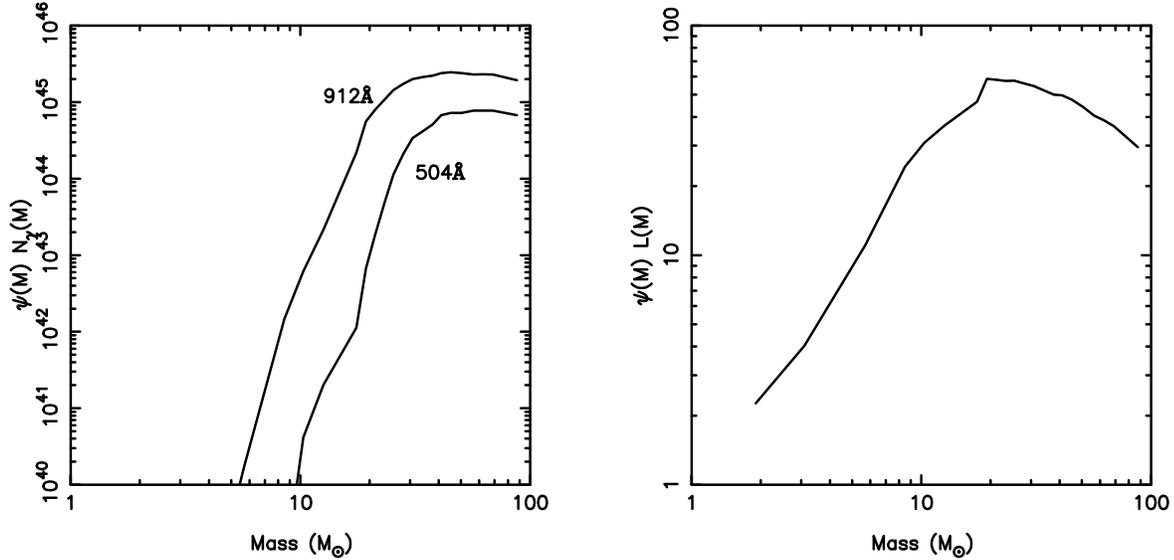
{\bf Figure 1:} The dependence of $\psi(M)\, n_\gamma(M)$ (left) and $\psi(M)\,
L(M)$ (right) as a function of stellar mass $M$ for the main sequence CoStar
models.  We have plotted curves representing both the helium and hydrogen
ionising continuum.  The data for masses $\ls19$\Msolar come from Puxley
(1988), using Kurucz (1979) models, since there are no CoStar models for this
range.  There is a clear discontinuity at the transition as seen in the right
hand panel, but the actual results with regard to the expected total luminosity
and ionising photon rate for a cluster of stars are not particularly affected by
this.
}\end{minipage}
\end{center}

\begin{center}
\begin{minipage}{6.5in}{
\psfig{file=Figure2.ps,width=6.5in,angle=0,clip=}}\end{minipage}

\begin{minipage}{\textwidth}{
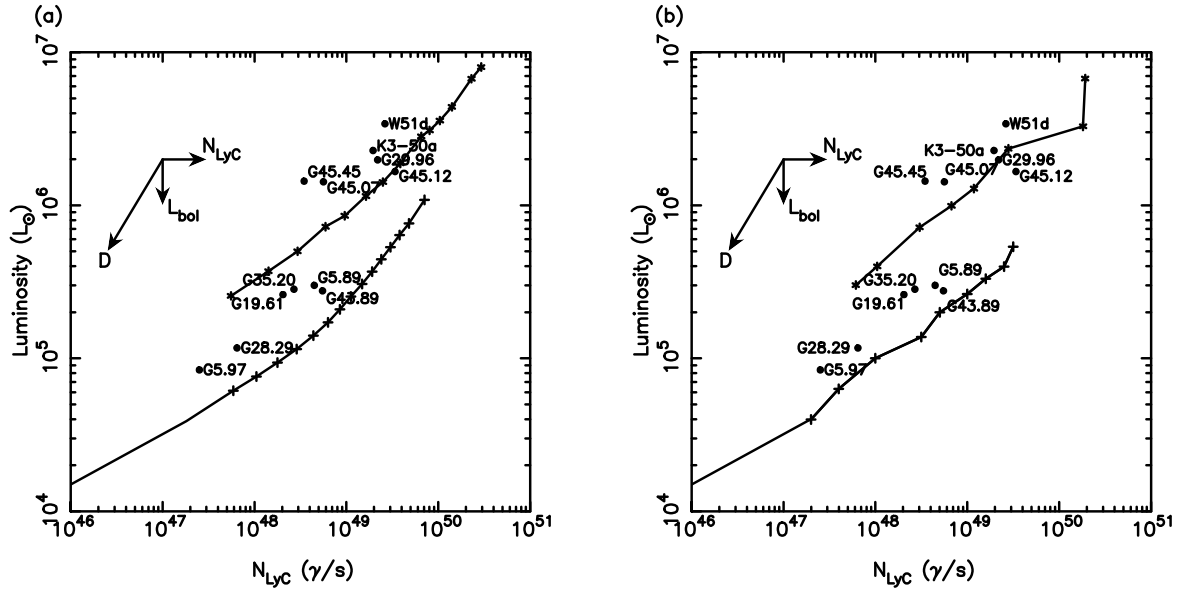
{\bf Figure 2:} The observed properties of our sample as taken from Table 2.
The model tracks represent single stars (lower) and clusters of stars (upper)
from (a) the CoStar model set and (b) the WM-basic model set.  The location of
the specific models given in Table 3 of Schaerer \& de Koter (1997) are marked
by $+$ in (a).  These span the range B0.5V to O3V.  Cluster models were also
constructed spanning this range.  These were normalised so that at any upper
mass cut-off, there was exactly one star of that spectral type.  The locations
of these cluster models are marked by $\ast$.  The solid line for the single
stars shows the extension to lower masses assuming the use of the Kurucz (1979)
models as described in the text.  The locations of the WM-basic models given in
Table 1 of Smith et al.\ (2002) are similarly indicated in (b).  Note that the
Smith et al.\ grid has coarser sampling than the CoStar set.  The arrows
indicate the magnitude and direction of the likely error in the observational
data.  }\end{minipage}
\end{center}

\begin{center}
\begin{minipage}{6.5in}{
\psfig{file=Figure3.ps,width=6.5in,angle=0,clip=}}\end{minipage}
\begin{minipage}{\textwidth}{
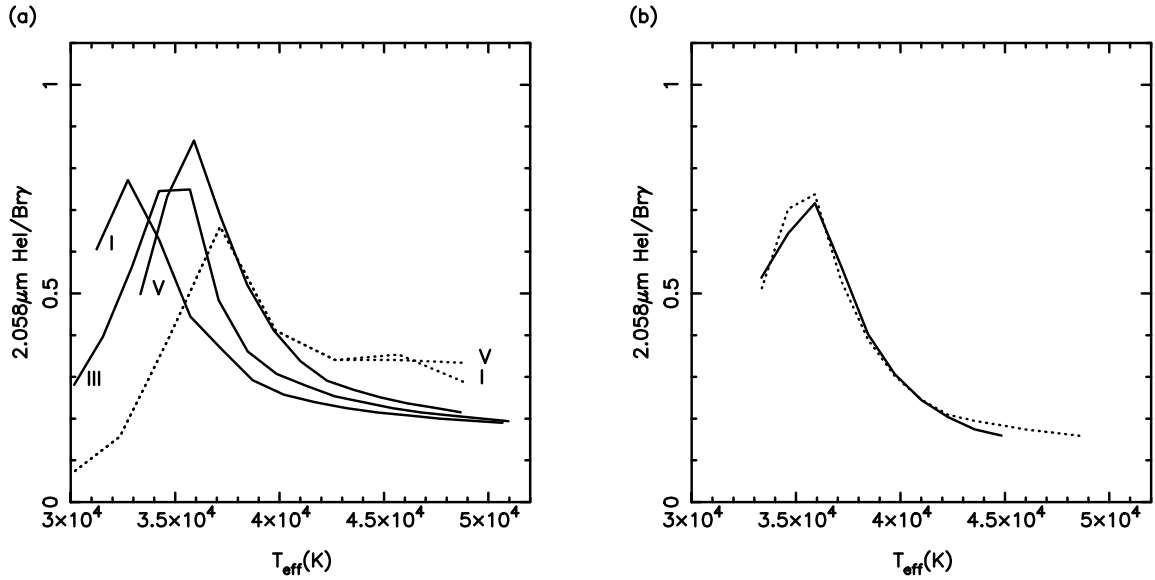
{\bf Figure 3:} Models of the HeI 2$^1$P--2$^1$S/HI Br$\gamma$ ratio as (a) a
function of the luminosity class of the exciting star, and of the input stellar
model set, and (b) a model stellar cluster.  In (a) the CoStar models are shown
as solid lines, and the WM-basic models as dotted lines.  Note that the
WM-basic grid is coarser than the CoStar grid, so the significance of the shift
in temperature of the peak in the line ratio between CoStar and WM-basic dwarf
models is not clear.  The larger line ratio at high effective temperature seen
with the WM-basic models may in part be due to the lower luminosity of the
dwarf models given in Smith et al.\ (2002).  In (b) the comparison is between
the results for the CoStar dwarf models when considered as a single star
(dotted line) and a cluster with standard IMF.  Clearly this line ratio depends
on the temperature of the hottest star present more than any other feature of
the IMF.  }\end{minipage}
\end{center}

\begin{center}
\begin{minipage}{6.5in}{
\psfig{file=Figure4.ps,width=6.5in,angle=0,clip=}}\end{minipage}
\begin{minipage}{\textwidth}{
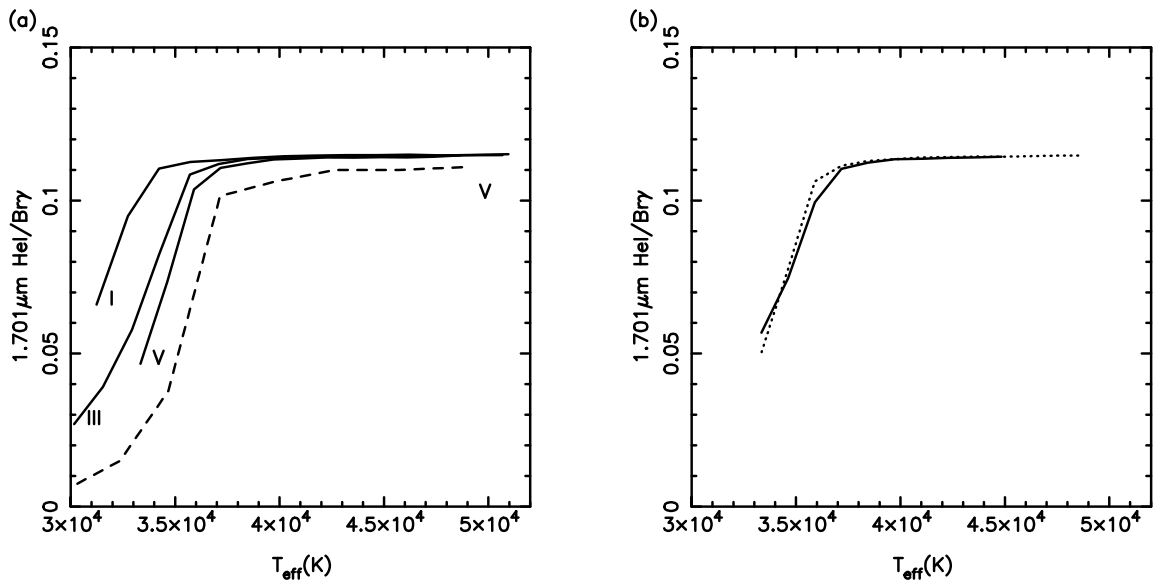
{\bf Figure 4:} The same models as shown in Figure 3 but for the HeI
4$^{3}$D--3$^{3}$P/HI Br$\gamma$ ratio.  Note the relatively small offset
between the CoStar and WM-basic dwarf models in (a), indicating that the
comparison of HeI/HI line ratios is not a good test of the detailed shape of
the spectral energy distribution of the exciting star(s).  Clearly (b) shows
there are no differences for this ratio between a single star and a cluster.
}\end{minipage}
\end{center}

\begin{center}
\begin{minipage}{6.5in}{
\psfig{file=Figure5.ps,width=6.5in,angle=0,clip=}}\end{minipage}
\begin{minipage}{\textwidth}{
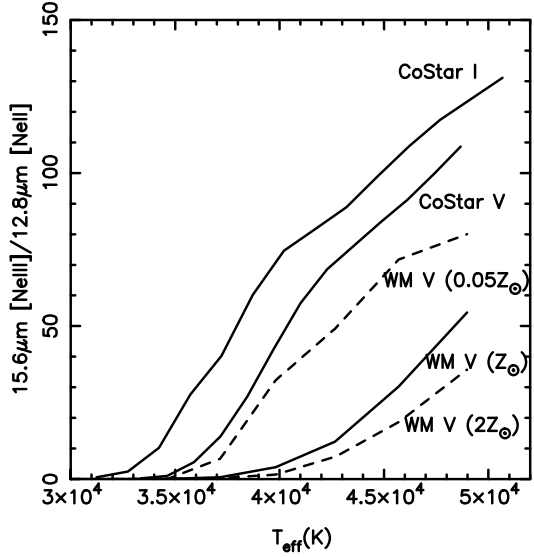
{\bf Figure 5:} A similar comparison to Figure 3 and 4 but using the
mid-infrared 15.6$\mu$m [NeIII] to 12.8$\mu$m [NeII] line ratio.  This plot
shows that there are significant detectable differences between the stellar
models with forbidden metal lines.  It also shows the role that metallicity
plays in determining such a line ratio, with a difference of a factor of two or
more in the final ratio.
}\end{minipage}
\end{center}

\begin{center}
\begin{minipage}{6.5in}{
\psfig{file=Figure6.ps,width=6.5in,angle=0,clip=}}\end{minipage}
\begin{minipage}{\textwidth}{
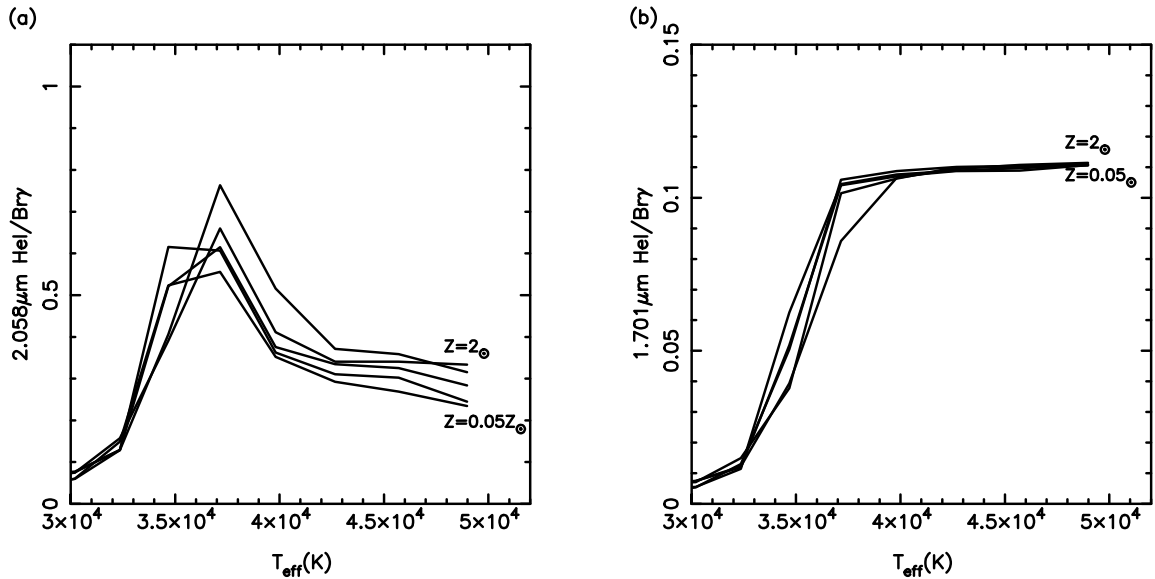
{\bf Figure 6:} The dependence of (a) the HeI 2$^1$P--2$^1$S/HI Br$\gamma$
ratio and (b) the HeI 4$^{3}$D--3$^{3}$P/HI Br$\gamma$ ratio as a function of
metallicity using the WM-basic model set.  Clearly, metallicity has a weak
effect on determining these line ratios.  }\end{minipage}
\end{center}

\begin{center}
\begin{minipage}{6.5in}{
\psfig{file=Figure7.ps,width=6.5in,angle=0,clip=}}\end{minipage}
\begin{minipage}{\textwidth}{
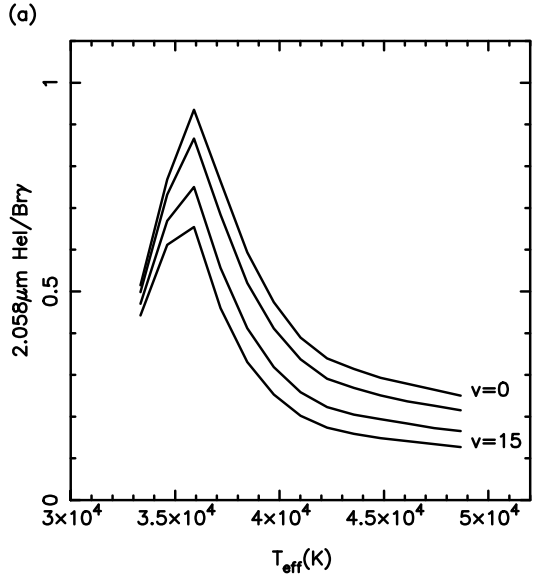
{\bf Figure 7:} Variation of the HeI 2$^1$P--2$^1$S/HI Br$\gamma$ ratio as a
function of turbulent velocity.  The models shown use CoStar dwarf models and
the turbulent velocity has values of 0, 5, 10 and 15kms$^{-1}$.  The result is
very similar to that obtained for planetary nebulae in Paper 1.
}\end{minipage}
\end{center}

\begin{center}
\begin{minipage}{6.5in}{
\psfig{file=Figure8.ps,width=6.5in,angle=0,clip=}}\end{minipage}
\begin{minipage}{\textwidth}{
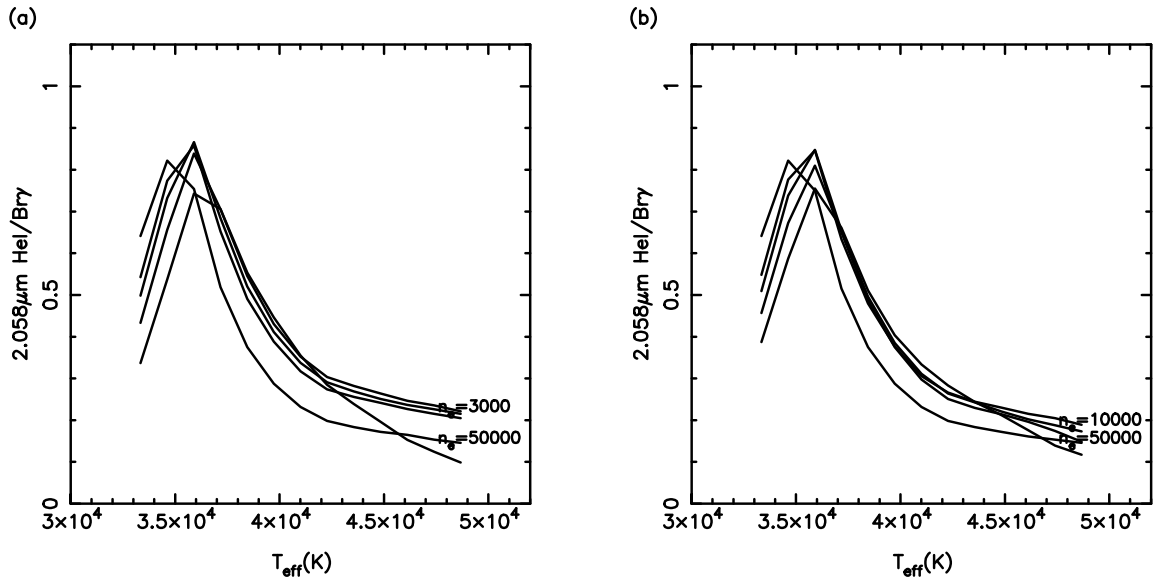
{\bf Figure 8:} Models of the HeI 2$^1$P--2$^1$S/HI Br$\gamma$ ratio for a
variety of electron densities.  In (a), the density is constant throughout the
HII region, with values of $n_e=$1000, 3000, 6000, 10000 and 50000cm$^{-3}$.
In (b), a constant density of $n_e=$1000cm$^{-3}$ is superposed on a Gaussian
radial distribution with peak density of $n_e=$1000, 3000, 6000, 10000 and
50000cm$^{-3}$.  The Gaussian has a full width at half maximum of 0.3~pc.  Some
of the specific models are indicated on the plot.  All models use CoStar dwarf
stellar atmospheres.  }\end{minipage}
\end{center}

\begin{center}
\begin{minipage}{6.5in}{
\psfig{file=Figure9.ps,width=6.5in,angle=0,clip=}}\end{minipage}
\begin{minipage}{\textwidth}{
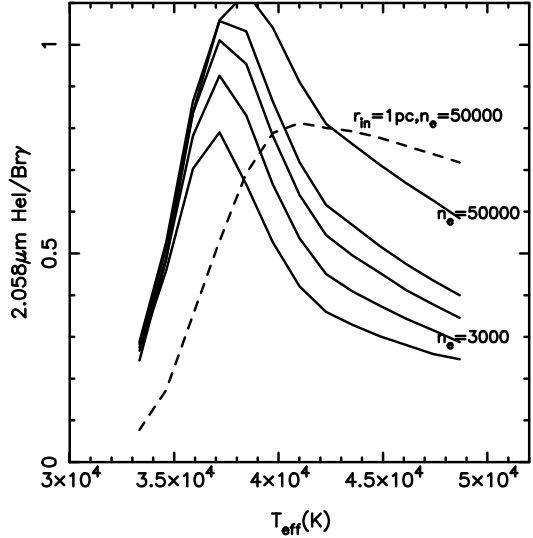
{\bf Figure 9:} Models of the HeI 2$^1$P--2$^1$S/HI Br$\gamma$ ratio for an
extended HII region excited by a single CoStar dwarf stellar atmosphere.  We
assume a flat electron density.  The solid lines show the same densities
as in Figure 8(a)
with an inner nebular radius of 0.3~pc.  The dashed line shows a single density
but with an inner radius of 1~pc.  These models correspond to a progressive
decline in the ionisation parameter at the inner surface of the nebula.  The
peak in the ratio clearly shifts to higher effective temperature, and the role
of collisional de-excitation of the 2$^3$S level must also increase with
decreasing ionisation parameters to explain these results.  }\end{minipage}
\end{center}

\begin{center}
\begin{minipage}{6.5in}{
\psfig{file=Figure10.ps,width=6.5in,angle=0,clip=}}\end{minipage}
\begin{minipage}{\textwidth}{
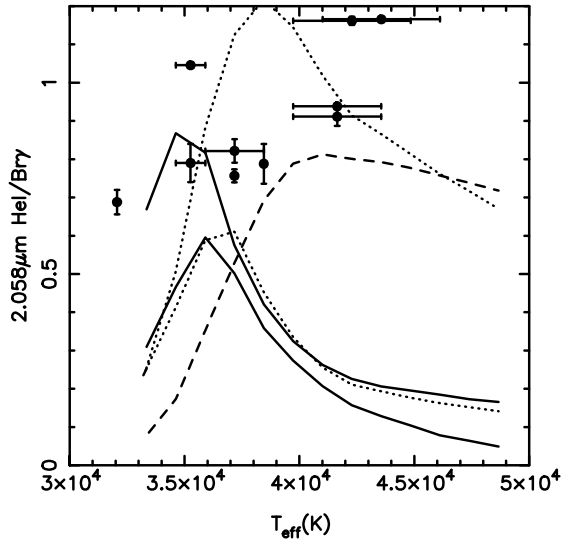
{\bf Figure 10:} Observed ratios of HeI 2$^1$P--2$^1$S and HI Br$\gamma$.
Vertical error bars are actual observed errors.  Horizontal error bars
represent the likely span in the effective temperature of the hottest star
present.  The extreme right point of the error bar represents $T_{eff}$ for a
single exciting star, the extreme left point represents $T_{eff}$ for a cluster
as outlined in Section 2.  Therefore the observed data point should be thought
of as lying at either extreme of the error bar, but not the central position.
Real observational error in our indirect spectral typing from Figure 2
can shift these points as much as 4000K.
The models are a selection of those from Figures 7, 8 and 9, showing the
range given by varying electron density, turbulent velocity and ionisation
parameters for the CoStar dwarf model set.  The solid lines show models with
$n_e=50000$cm$^{-3}$ and turbulent velocity of 0kms$^{-1}$ (upper curves) and
$n_e=1000$cm$^{-3}$ and turbulent velocity of 15kms$^{-1}$ (lower curves).  The
dotted lines show the same models but with inner nebular radius of 0.3~pc.  The
dashed line shows a model with inner nebular radius of 1~pc,
$n_e=50000$cm$^{-3}$ and turbulent velocity of 5kms$^{-1}$.  The observed
ratio is not a good fit to any of these curves, though the models in which
collisional de-excitation dominates photoionisation of HeI 2$^3$S are
a better match.}
\end{minipage} \end{center}

\begin{center}
\begin{minipage}{6.5in}{
\psfig{file=Figure11.ps,width=6.5in,angle=0,clip=}}\end{minipage}
\begin{minipage}{\textwidth}{
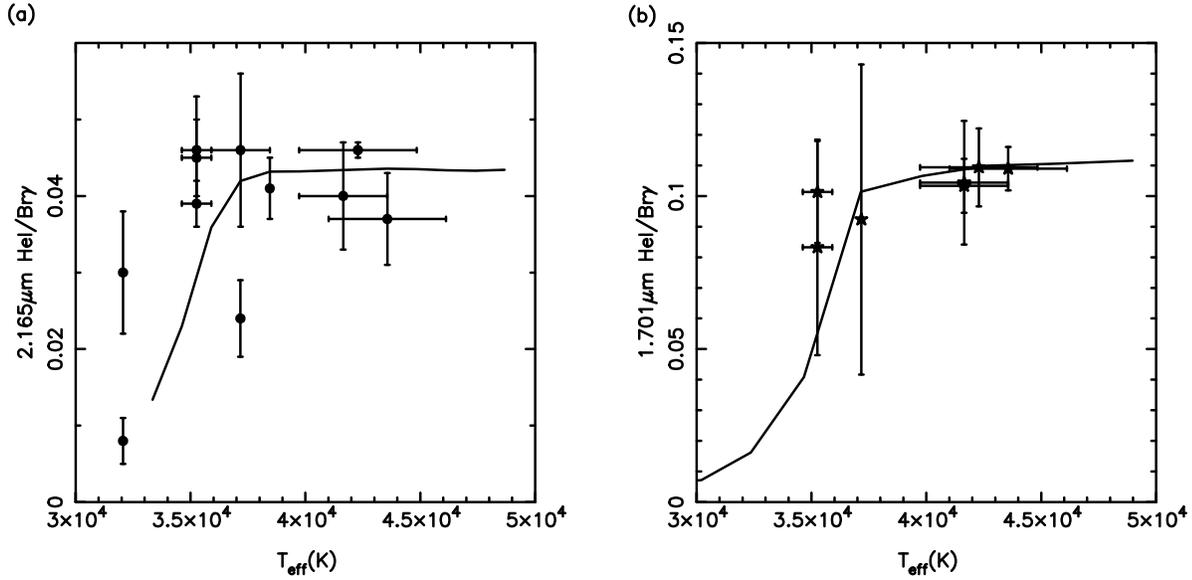
{\bf Figure 11:} Observed ratios of (a) HeI 7$^{3,1}$G--4$^{3,1}$F and (b) HeI
4$^{3}$D--3$^{3}$P with HI Br$\gamma$.  Vertical error bars are again actual
observed errors, and horizontal error bars have the same meaning as in Figure
10.  The models shown in (a) are for CoStar dwarfs, and in (b) for WM-basic
dwarfs.  There is essentially no variation of these curves with density,
turbulent velocity or radial density profile.  The observed data are mostly a
reasonable match to both CoStar and WM-basic dwarf models.  
The match argues that our indirect spectral typing is also
reasonable.
}\end{minipage}
\end{center}

\begin{center}
\begin{minipage}{6.5in}{
\psfig{file=Figure12.ps,width=6.5in,angle=0,clip=}}\end{minipage}

\begin{minipage}{\textwidth}{
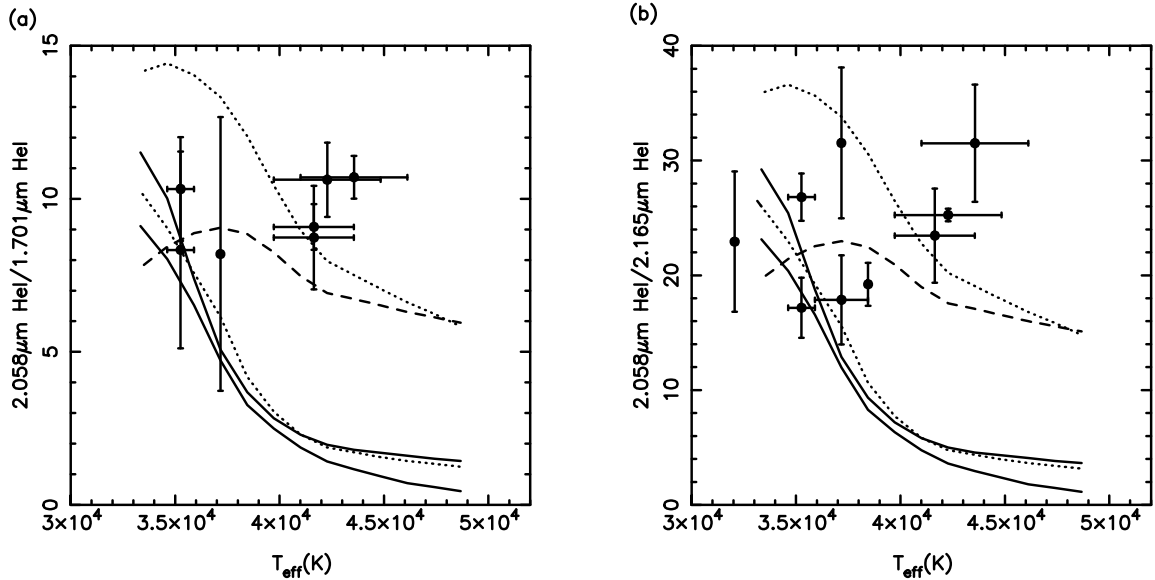
{\bf Figure 12:} Observed ratios of HeI 2$^1$P--2$^1$S with (a) HeI
4$^{3}$D--3$^{3}$P and (b) HeI 7$^{3,1}$G--4$^{3,1}$F.  The models plotted are
the same as in Figure 10, and the error bars have the same meaning as there.
Again high density low ionisation parameter models are the best match.
}\end{minipage}
\end{center}

\end{document}